\documentclass[onecolumn,showpacs,preprintnumbers,amsmath,amssymb]{revtex4}
\usepackage{amsmath}
\usepackage{amssymb}
\usepackage{epsfig}
\usepackage{graphicx}
\begin{document}
\def\pslash{\rlap{\hspace{0.02cm}/}{p}}
\def\eslash{\rlap{\hspace{0.02cm}/}{e}}
\title{The flavor-changing top-charm associated productions at ILC in littlest
       Higgs model with T parity }
\author{Yanju Zhang}
\author{Gongru Lu}
\author{Xuelei Wang}
\affiliation{College of Physics and Information Engineering, Henan
Normal University, Xinxiang, Henan 453007, P.R. China}
\date{\today}
\begin{abstract}
The littlest Higgs model with T-parity (LHT) has new
flavor-changing (FC) couplings with the Standard Model (SM) quarks,
which do not suffer strong constraints from electroweak precision
data. So these FC interactions may enhance the cross sections of
some flavor-changing neutral-current (FCNC) processes. In this work,
we study the FC top-charm associated productions via $e^{-}\gamma$
collision at the ILC. We find that the cross sections are
sensitive to the mirror quark masses. With reasonable values of the
parameters, the cross sections may reach the detectable level and
provide useful information about the relevant parameters in the LHT
model, especially setting an upper limit on the mirror quark masses.
\end{abstract}


\maketitle


\section{ Introduction}
An interesting solution to the hierarchy problem of the
Standard Model (SM) is the little Higgs theory \cite{little Higgs}.
In this theory the Higgs boson is regarded as a pseudo-Goldstone boson (PGB)
which can be naturally "little" in the current reincarnation of the PGB idea
called collective symmetry breaking.  Through such a collective symmetry breaking
mechanism, the one loop quadratic divergences in the Higgs boson mass can be avoided.
The littlest Higgs model (LH) \cite{LH} is the most
economical implementation of the little Higgs idea, which, however,
suffers strong constraints from electroweak precision
data \cite{constraints} due to the tree level mixing of heavy and
light mass eigenstates. So the LH model would require raising the
mass scale and thus reintroduce the fine-tuning in the Higgs
potential \cite{fine-tuning}. To solve this problem, a $Z_2$ discrete
symmetry called T-parity is introduced \cite{LHT}. Under this T-parity
the SM particles are even while most of the new particles at the TeV scale
are odd. T-parity explicitly forbids any tree-level contribution from
the heavy gauge bosons to the observables involving only SM
particles as external states. Since in the LHT model
the corrections to the precision electroweak observables are generated
at loop-level and suppressed, the fine tuning can be avoided \cite{scale}.

It is well known that the flavor-changing neutral-current (FCNC)
interactions are absent at tree level and extremely small at loop
levels in the SM due to the GIM mechanism. However, in the LHT
model, the flavor-changing (FC) interactions between the SM fermions
and the mirror fermions, which are parameterized by the newly
CKM-like unitary mixing matrices, may have significant contributions
to some FC processes. So much attention was paid on the FC interactions
in LHT model in recent years. Firstly, the LHT flavor structure was
analysed and some constraints on the mirror fermion mass spectrum
was obtained from an one-loop analysis of neutral meson mixing in the
$K,B$ and $D$ systems \cite{FC-LHT3}. Then an extensive study of FC
transitions in the LHT model was performed in \cite{FC-LHT2,FC-LHT1,FC-LHT6},
which considered all prominent rare $K$ and $B$ decays and presented a collection
of Feynman rules to the order of $v^2/f^2$. Motivated by the experimental
evidence of meson oscillations in the $D$-system,
the impact of $D^0-\bar{D}^0$ mixing on the LHT flavor structure
was investigated in \cite{FC-LHT5}.
Furthermore, the LHT flavor study was extended to the lepton flavor
violating decays in \cite{FC-LHT4}.

The International Linear Collider(ILC) with the center of mass
(c.m.) energy from 200 GeV to 1.0 TeV and high luminosity has been
proposed \cite{ILC}. Due to its rather clean environment and high
luminosity, the ILC will be an ideal machine for probing new
physics. In such a collider, in addition to $e^{+}e^{-}$ collision,
we may also realize $\gamma\gamma$ or $e^{-}\gamma$ collision with
the photon beams generated by the backward Compton scattering of
incident electron- and laser-beams \cite{er}. In particular, as the
heaviest fermion with a mass of the order of the electroweak scale,
the top quark is naturally regarded to be more sensitive to new
physics than other fermions. Therefore the top quark FCNC processes
at the ILC would provide an important test for new physics. This
stimulates many attempts in probing new physics via rare top quark
decays \cite{Rare top decay} or FC production processes at ILC
\cite{eerrtc-MSSM,rrtc-MSSM,eetc-2HDM,eerrtc-TC2,eetc-TC2,rrtc-2HDM,
rrtc-TC2,tc-effective-LC1,tc-effective-LC2,tc-effective-LC3,wang}.
The FC couplings between the SM fermions and the mirror fermions can
also induce the loop-level $tcV(V=\gamma,Z,g)$ couplings in the LHT
model. Studies \cite{tcv-LHT,tc-LHT,et-LHT} showed that some
processes induced by such $tcV$ couplings in the LHT model can be
significantly enhanced. In this paper, we will study the process
$e^-\gamma\rightarrow e^-t\bar{c}$ induced by the $tcV$ couplings in
the LHT model and compare with the process
$e^+e^-(\gamma\gamma)\rightarrow t\bar{c}$ studied
previously\cite{tc-LHT}. Note that these processes have been studied
thoroughly in other models, such as the MSSM
\cite{eerrtc-MSSM,rrtc-MSSM}, the 2HDM\cite{eetc-2HDM,rrtc-2HDM} and
the TC2 model\cite{eerrtc-TC2,eetc-TC2,rrtc-TC2}, and also in the
model-independent way
\cite{tc-effective-LC1,tc-effective-LC2,tc-effective-LC3}. They
showed that the production rates of such processes could be
significantly enhanced by several orders compared to the SM
predictions \cite{eerrtc-MSSM,tc-SM}. As found in other new physics
models \cite{eerrtc-MSSM,eerrtc-TC2}, the process
$e^-\gamma\rightarrow e^-t\bar{c}$ has a much larger rate than
$e^+e^-\rightarrow t\bar{c}$ for some part of the parameter space.
In our study we will compare the LHT prediction with those predicted
by other new physics models. Such an analysis will help to
distinguish different models once the measurements are observed at
the ILC.

This paper is organized as follows. In Sec.II, we briefly review the
LHT model. In Sec.III, we present the detailed calculations for the
production processes. The numerical results of the production cross
sections and discussions are shown in Sec.IV. Our conclusions are
listed in the last section.

\section{The littlest Higgs model with T-parity}

The LHT model\cite{LHT} is based on a non-linear $\sigma$ model
describing an  $SU(5)/SO(5)$ symmetry breaking with a locally gauged
sub-group $[SU(2)\times U(1)]^2$. The $SU(5)$ symmetry spontaneously
breaks down to $SO(5)$ at the scale $f\sim\mathcal {O}(TeV)$.  From
the $SU(5)/SO(5)$ breaking, there arise 14 Nambu-Goldstone bosons
which are described by the matrix $\Pi$, given explicitly by
\begin {equation}
\Pi=
\begin{pmatrix}
-\frac{\omega^0}{2}-\frac{\eta}{\sqrt{20}}&-\frac{\omega^+}{\sqrt{2}}
&-i\frac{\pi^+}{\sqrt{2}}&-i\phi^{++}&-i\frac{\phi^+}{\sqrt{2}}\\
-\frac{\omega^-}{\sqrt{2}}&\frac{\omega^0}{2}-\frac{\eta}{\sqrt{20}}
&\frac{v+h+i\pi^0}{2}&-i\frac{\phi^+}{\sqrt{2}}&\frac{-i\phi^0+\phi^P}{\sqrt{2}}\\
i\frac{\pi^-}{\sqrt{2}}&\frac{v+h-i\pi^0}{2}&\sqrt{4/5}\eta&-i\frac{\pi^+}{\sqrt{2}}&
\frac{v+h+i\pi^0}{2}\\
i\phi^{--}&i\frac{\phi^-}{\sqrt{2}}&i\frac{\pi^-}{\sqrt{2}}&
-\frac{\omega^0}{2}-\frac{\eta}{\sqrt{20}}&-\frac{\omega^-}{\sqrt{2}}\\
i\frac{\phi^-}{\sqrt{2}}&\frac{i\phi^0+\phi^P}{\sqrt{2}}&\frac{v+h-i\pi^0}{2}&-\frac{\omega^+}{\sqrt{2}}&
\frac{\omega^0}{2}-\frac{\eta}{\sqrt{20}}
\end{pmatrix}
\end{equation}
Here, $H=(-i\pi^+\sqrt{2},(v+h+i\pi^0)/2)^T$ is the SM Higgs doublet
and $\Phi$ is a physical scalar triplet with
\begin {equation}
\Phi=
\begin{pmatrix}
-i\phi^{++}&-i\frac{\phi^+}{\sqrt{2}}\\
-i\frac{\phi^+}{\sqrt{2}}&\frac{-i\phi^0+\phi^P}{\sqrt{2}}
\end{pmatrix}
\end{equation}

In the LHT model, a T-parity discrete symmetry is introduced to make
the model consistent with the electroweak precision data. Under the
T-parity, the fields $\Phi,\omega$ and $\eta$ are odd, and the SM
Higgs doublet $H$ is even.

For the gauge subgroup $[SU(2)\times U(1)]^2$ of the global symmetry
$SU(5)$, from the first step of symmetry breaking $[SU(2)\times
U(1)]^2\rightarrow SU(2)_L\times U(1)_Y$, which is identified as the
SM electroweak gauge group, the Goldstone bosons
$\omega^{0},\omega^{\pm}$ and $\eta$ are respectively eaten by the
new T-odd gauge bosons$Z_{H},W_{H}$ and $A_{H}$, which obtain masses
at the order of $~\mathcal {O}(v^2/f^2)$
\begin{eqnarray}
M_{Z_H}\equiv
M_{W_H}=fg(1-\frac{v^2}{8f^2}),~~~~M_{A_H}=\frac{fg'}{\sqrt{5}}(1-\frac{5v^2}{8f^2}),
\end{eqnarray}
with $g,g'$ being the corresponding coupling constants of $SU(2)_L$
and $U(1)_Y$.

From the second step of symmetry breaking $SU(2)_L\times
U(1)_Y\rightarrow U(1)_{em}$, the masses of the SM T-even gauge
bosons $Z$ and $W$ are generatec through eating the Goldstone bosons
$\pi^{0}$ and $\pi^{\pm}$. They are given at $~\mathcal
{O}(v^2/f^2)$ by
\begin{eqnarray}
M_{W_L}=\frac{gv}{2}(1-\frac{v^2}{12f^2}),
~~M_{Z_L}=\frac{gv}{2cos\theta_W}(1-\frac{v^2}{12f^2}),~~ M_{A_L}=0.
\end{eqnarray}

A consistent and phenomenologically viable implementation of
T-parity in the fermion sector requires the introduction of mirror
fermions. The T-even fermion section consists of the SM quarks,
leptons and an additional heavy quark $T_+$. The T-odd fermion
sector consists of three generations of mirror quarks and leptons
and an additional heavy quark $T_-$. Only the mirror quarks
$(u^i_H,d^i_H)$ are involved in this paper. The mirror fermions get
masses
\begin{eqnarray}
m^u_{H_i}=\sqrt{2}\kappa_if(1-\frac{v^2}{8f^2})\equiv
m_{H_i}(1-\frac{v^2}{8f^2}), \\
\nonumber
m^d_{H_i}=\sqrt{2}\kappa_if\equiv m_{H_i},
\end{eqnarray}
where the Yukawa couplings $\kappa_i$ can in general depend on the
fermion species $i$.

The mirror fermions induce a new flavor structure and there are
four CKM-like unitary mixing matrices in the mirror fermion
sector:
\begin{eqnarray}
V_{H_u},~~V_{H_d},~~V_{H_l},~~V_{H_{\nu}}.
\end{eqnarray}
These mirror mixing matrices are involved in the FC interactions
between the SM fermions and the T-odd mirror fermions which are
mediated by the T-odd heavy gauge bosons or the Goldstone bosons.
$V_{H_u}$ and $V_{H_d}$ satisfy the relation
\begin{eqnarray}
V^{\dag}_{H_u}V_{H_d}=V_{CKM}.
\end{eqnarray}
We parameterize the $V_{H_d}$ with three angles
$\theta^d_{12},\theta^d_{23},\theta^d_{13}$ and three phases
$\delta^d_{12},\delta^d_{23},\delta^d_{13}$

\begin {eqnarray}
V_{H_d}=
\begin{pmatrix}
c^d_{12}c^d_{13}&s^d_{12}c^d_{13}e^{-i\delta^d_{12}}&s^d_{13}e^{-i\delta^d_{13}}\\
-s^d_{12}c^d_{23}e^{i\delta^d_{12}}-c^d_{12}s^d_{23}s^d_{13}e^{i(\delta^d_{13}-\delta^d_{23})}&
c^d_{12}c^d_{23}-s^d_{12}s^d_{23}s^d_{13}e^{i(\delta^d_{13}-\delta^d_{12}-\delta^d_{23})}&
s^d_{23}c^d_{13}e^{-i\delta^d_{23}}\\
s^d_{12}s^d_{23}e^{i(\delta^d_{12}+\delta^d_{23})}-c^d_{12}c^d_{23}s^d_{13}e^{i\delta^d_{13}}&
-c^d_{12}s^d_{23}e^{i\delta^d_{23}}-s^d_{12}c^d_{23}s^d_{13}e^{i(\delta^d_{13}-\delta^d_{12})}&
c^d_{23}c^d_{13}
\end{pmatrix}
\end{eqnarray}
The matrix $V_{H_u}$ is then determined through
$V_{H_u}=V_{H_d}V^{\dag}_{CKM}$. As in the case of the CKM matrix
the angles $\theta^d_{ij}$ can all be made to lie in the first
quadrant with $0\leq
\delta^d_{12},\delta^d_{23},\delta^d_{13}<2\pi$.
\section{Top-charm quark associated productions in the LHT model}
In the LHT model, there are FC interactions between SM quarks and
T-odd mirror quarks which are mediated by the heavy T-odd gauge
bosons or Goldstone bosons. With these FC couplings, the loop-level
FC couplings $t\bar{c}\gamma(Z)$  can be induced and the relevant
Feynman diagrams are shown in Fig.1.

The effective one-loop level couplings $t\bar{c}\gamma(Z)$ can be
directly calculated by the method introduced in Ref.\cite{method}.
The relevant Feynman rules can be found in Ref.\cite{FC-LHT1}. We
list the explicit forms of
$\Gamma^{\mu}_{t\bar{c}\gamma}(p_t,p_{\bar{c}})$ and
$\Gamma^{\mu}_{t\bar{c}Z}(p_t,p_{\bar{c}})$ in Appendix.

\begin{figure}[h]
\includegraphics {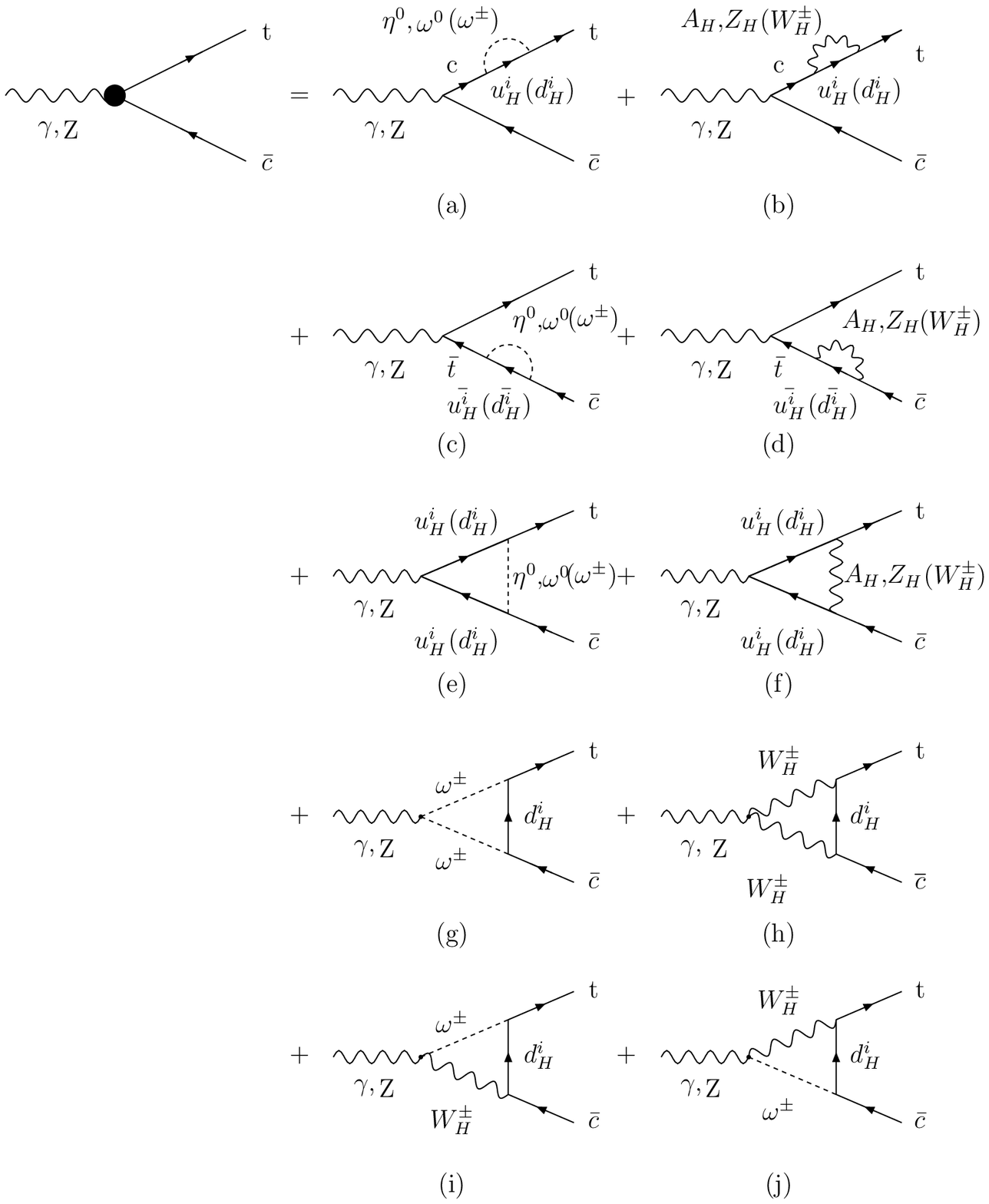}
\caption{\small One-loop contributions of the LHT model to the
couplings $t\bar{c}\gamma(Z)$.}
\end{figure}

The FC couplings $t\bar{c}\gamma(Z)$ can contribute to the top-charm
associated productions. Here we reconsider the processes
$e^+e^-\rightarrow t\bar{c}$ in $e^{+}e^{-}$ collision and
$\gamma\gamma\rightarrow t\bar{c}$ in $\gamma\gamma$ collision which
have been calculated in the literature\cite{tc-LHT}, since we had a
mistake in the program calculation of them before. So we will no
longer present the relevant Feynman diagrams and the corresponding
production amplitudes of the processes in the following.

\begin{figure}
\includegraphics {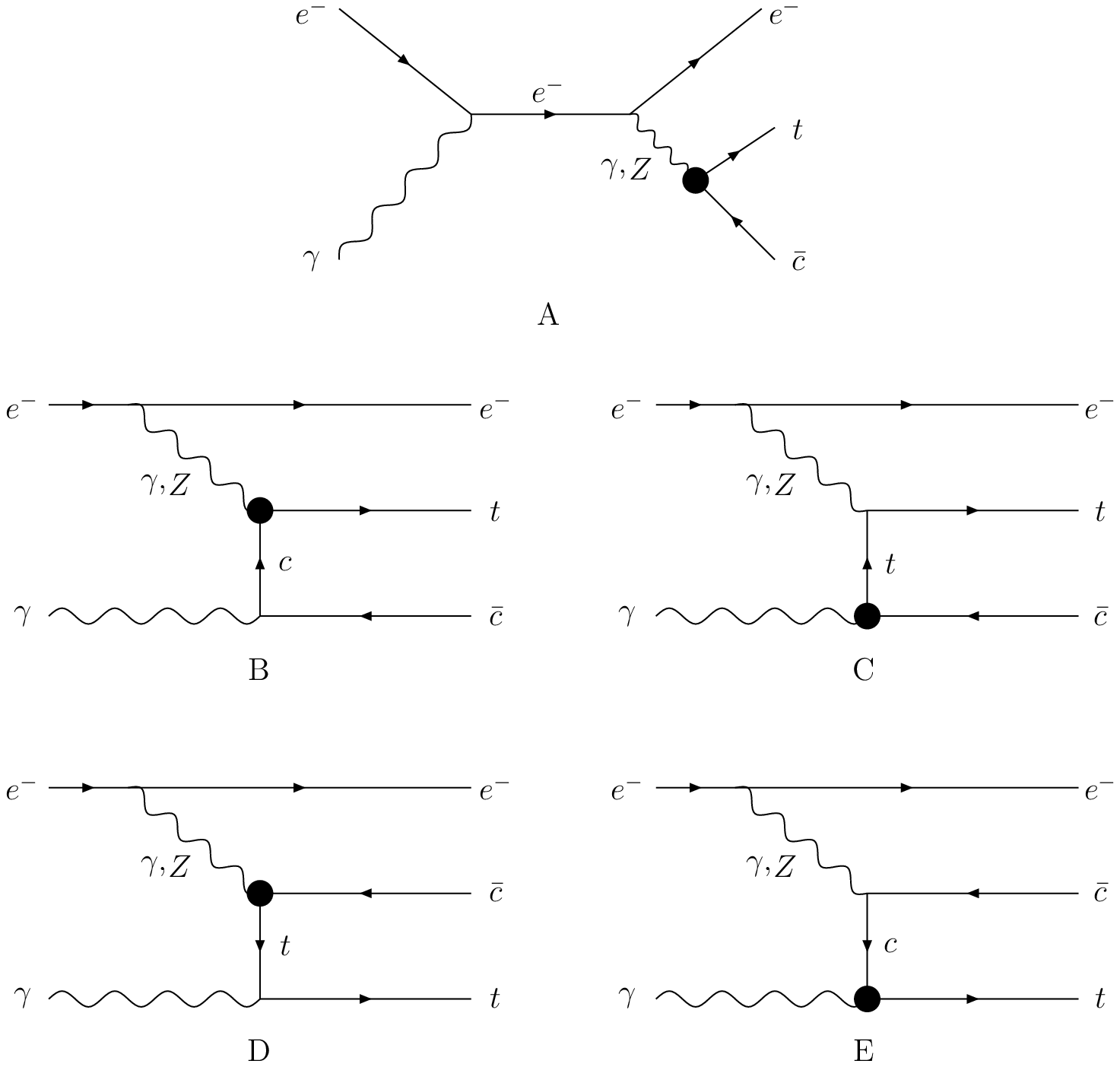}
\caption{\small The Feynman diagrams for $e^{-}\gamma\rightarrow
e^{-}t\bar{c}$ in the LHT model.} \label{fig:fig2}
\end{figure}
We also focus on the process $e^{-}\gamma\rightarrow e^{-}t\bar{c}$
in $e^{-}\gamma$ collision. The $t\bar{c}$ production in
$e^{-}\gamma$ collision proceeds through the process
$e^-\gamma\rightarrow e^{-}\gamma^{\ast}(Z^{\ast})\gamma\rightarrow
e^{-}t\bar{c}$, where the $\gamma$-beam is generated by the backward
Compton scattering of incident electron- and laser-beam and the
$\gamma^{\ast}(Z^{\ast})$ is radiated from $e^{-}$ beam. The
corresponding Feynman diagram is shown in Fig.2(A-E) and the
invariant production amplitudes of the process can be written as
\begin{eqnarray}
M_A^{\gamma}&=&-e^{2}G(p_1+p_2,0)G(p_3+p_4,0)\bar{u}_{e^{-}}(p_{5})\gamma_{\mu}(\pslash_1+\pslash_2)
\rlap/\epsilon(p_2)u_{e^{-}}(p_{1})~~~~~~~~~~~~~~~~~~~\nonumber \\
&&\times\bar{u}_{t}(p_{3})\Gamma^{\mu}_{t\bar{c}\gamma}(p_{3},p_4)v_{\bar{c}}(p_{4}),
 \end{eqnarray}
\begin{eqnarray}
M_{A}^{Z}&=&\frac{eg}{\cos\theta_{W}}G(p_1+p_2)G(p_3+p_4,M_{Z})\bar{u}_{e^{-}}(p_{5})\gamma_{\mu}
[(-\frac{1}{2}+\sin^{2}\theta_{W})P_{L}~~~~~~~~~~~~~~~~~\nonumber \\
&&+(\sin^{2}\theta_{W})P_{R}](\pslash_1+\pslash_2)\rlap/\epsilon(p_2)u_{e^{-}}(p_{1})
\bar{u}_{t}(p_{3})\Gamma^{\mu}_{t\bar{c}Z}(p_{3},p_4)v_{\bar{c}}(p_{4}),
\end{eqnarray}

\begin{eqnarray}
M_B^{\gamma}&=&\frac{2e^{2}}{3}G(p_1-p_5,0)G(p_2-p_4,m_{c})\bar{u}_{e^{-}}(p_{5})\gamma_{\mu}u_{e^{-}}(p_{1})
~~~~~~~~~~~~~~~~~~~~~~~~~~~~~~~~~\nonumber \\
&&\times\bar{u}_{t}(p_{3})\Gamma^{\mu}_{t\bar{c}\gamma}(p_{3},p_4-p_2)
(\pslash_2-\pslash_4+m_{c})\rlap/\epsilon(p_2)v_{\bar{c}}(p_{4}),
\end{eqnarray}
\begin{eqnarray}
M_B^{Z}&=&-\frac{2eg}{3\cos\theta_{W}}G(p_1-p_5,M_{Z})G(p_2-p_4,m_{c})\bar{u}_{e^{-}}(p_{5})\gamma_{\mu}
[(-\frac{1}{2}+\sin^{2}\theta_{W})P_{L}~~~~~~\nonumber\\
&&+(\sin^{2}\theta_{W})P_{R}]u_{e^{-}}(p_{1})
\bar{u}_{t}(p_{3})\Gamma^{\mu}_{t\bar{c}Z}(p_{3},p_4-p_2)
(\pslash_2-\pslash_4+m_{c})\rlap/\epsilon(p_2)v_{\bar{c}}(p_{4}),
\end{eqnarray}

\begin{eqnarray}
M_C^{\gamma}&=&\frac{2e^{2}}{3}G(p_1-p_5,0)G(p_2-p_4,m_{t})\bar{u}_{e^{-}}(p_{5})\gamma_{\mu}u_{e^{-}}(p_{1})
~~~~~~~~~~~~~~~~~~~~~~~~~~~~~~~~~\nonumber \\
&&\times\bar{u}_{t}(p_{3})\gamma^{\mu}(\pslash_2-\pslash_4+m_{t})\Gamma^{\nu}_{t\bar{c}\gamma}(p_2-p_4,p_{4})
\epsilon_{\nu}(p_2)v_{\bar{c}}(p_{4}),
\end{eqnarray}
\begin{eqnarray}
M_C^{Z}&=&-\frac{g^{2}}{\cos^{2}\theta_{W}}G(p_1-p_5,M_{Z})G(p_2-p_4,m_{t})\bar{u}_{e^{-}}(p_{5})
\gamma_{\mu}[(-\frac{1}{2}+\sin^{2}\theta_{W})P_{L}
~~~~~~~~\nonumber\\
&&+(\sin^{2}\theta_{W})P_{R}]u_{e^{-}}(p_{1})
\bar{u}_{t}(p_{3})\gamma^{\mu}[(\frac{1}{2}-\frac{2}{3}\sin^{2}\theta_{W})P_{L}
-\frac{2}{3}(\sin^{2}\theta_{W})P_{R}]~~~~~~\nonumber\\
&&\times(\pslash_2-\pslash_4+m_{t})\Gamma^{\nu}_{t\bar{c}\gamma}(p_2-p_4,p_{4})
\epsilon_{\nu}(p_2)v_{\bar{c}}(p_{4}),
\end{eqnarray}

\begin{eqnarray}
M_D^{\gamma}&=&\frac{2e^{2}}{3}G(p_1-p_5,0)G(p_3-p_2,m_{t})\bar{u}_{e^{-}}(p_{5})\gamma_{\mu}u_{e^{-}}(p_{1})
~~~~~~~~~~~~~~~~~~~~~~~~~~~~~~~~~\nonumber \\
&&\times\bar{u}_{t}(p_{3})\rlap/\epsilon(p_2)(\pslash_3-\pslash_2+m_{t})\Gamma^{\mu}_{t\bar{c}\gamma}(p_3-p_2,p_{4})
v_{\bar{c}}(p_{4}),
\end{eqnarray}
\begin{eqnarray}
M_D^{Z}&=&-\frac{2eg}{3\cos\theta_{W}}G(p_1-p_5,M_{Z})G(p_3-p_2,m_{t})\bar{u}_{e^{-}}(p_{5})\gamma_{\mu}
[(-\frac{1}{2}+\sin^{2}\theta_{W})P_{L}~~~~~~\nonumber\\
&&+(\sin^{2}\theta_{W})P_{R}]u_{e^{-}}(p_{1})\bar{u}_{t}(p_{3})\rlap/\epsilon(p_2)
(\pslash_3-\pslash_2+m_{t})\Gamma^{\mu}_{t\bar{c}Z}(p_3-p_2,p_{4})v_{\bar{c}}(p_{4}),
\end{eqnarray}

\begin{eqnarray}
M_E^{\gamma}&=&\frac{2e^{2}}{3}G(p_1-p_5,0)G(p_3-p_2,m_{c})\bar{u}_{e^{-}}(p_{5})\gamma_{\mu}u_{e^{-}}(p_{1})
~~~~~~~~~~~~~~~~~~~~~~~~~~~~~~~~~\nonumber \\
&&\times\bar{u}_{t}(p_{3})\Gamma^{\nu}_{t\bar{c}\gamma}(p_3,p_2-p_3)\epsilon_{\nu}(p_2)(\pslash_3-\pslash_2+m_{c})
\gamma^{\mu}v_{\bar{c}}(p_{4}),
\end{eqnarray}
\begin{eqnarray}
M_E^{Z}&=&-\frac{g^{2}}{\cos^{2}{\theta}_{W}}G(p_1-p_5,M_{Z})G(p_3-p_2,m_{c})\bar{u}_{e^{-}}(p_{5})\gamma_{\mu}
[(-\frac{1}{2}+\sin^{2}\theta_{W})P_{L}~~~~~~\nonumber\\
&&+(\sin^{2}\theta_{W})P_{R}]u_{e^{-}}(p_{1})
\bar{u}_{t}(p_{3})\Gamma^{\nu}_{t\bar{c}\gamma}(p_3,p_2-p_3)\epsilon_{\nu}(p_2)(\pslash_3-\pslash_2+m_{c})
\gamma^{\mu}~~~~~~\nonumber\\&&\times[(\frac{1}{2}-\frac{2}{3}\sin^{2}\theta_{W})P_{L}
-\frac{2}{3}(\sin^{2}\theta_{W})P_{R}] v_{\bar{c}}(p_{4}),
\end{eqnarray}
where $P_L=\frac{1}{2}(1-\gamma_5)$ and
$P_R=\frac{1}{2}(1+\gamma_5)$ are the left and right chirality
projectors. $p_{1},p_{2}$ are the momenta of the incoming
$e^-,\gamma$, and $p_{3},p_{4},p_{5}$ are the momenta of the
outgoing final states top quark, anti-charm quark and electron,
respectively. We also define $G(p, m)$ as $\frac{1}{p^2-m^2}$.

With the above amplitudes, we can directly obtain the production
cross section $\hat{\sigma}(\hat{s})$ for the subprocess
$e^{-}\gamma\rightarrow e^{-}t\bar{c}$ and the total cross sections
at the $e^+e^-$ linear collider can be obtained by folding
$\hat{\sigma}(\hat{s})$ with the photon distribution function $F(x)$ \cite{distribution}:
\begin{eqnarray}
\sigma_{e^{-}\gamma\rightarrow
e^{-}t\bar{c}}(s_{e^{+}e^{-}})=\int^{x_{max}}_{(m_{t}+m_{c})^{2}/s_{e^{+}e^{-}}}
dxF(x)\hat{\sigma}(\hat{s})
\end{eqnarray}
where $s$ is the c.m. energy squared for $e^+e^-$. The subprocess
occurs effectively at $\hat{s}=xs$, and $x$ is the fractions of the
electron energies carried by the photons. The explicit form of the
photon distribution function $F(x)$ is
\begin{eqnarray}
\displaystyle F(x)=\frac{1}{D(\xi)}\left[1-x+\frac{1}{1-x}
-\frac{4x}{\xi(1-x)}+\frac{4x^2}{\xi^2(1-x)^2}\right],
\end{eqnarray}
with
\begin{eqnarray}
\displaystyle D(\xi)=\left(1-\frac{4}{\xi}-\frac{8}{\xi^2}\right)
\ln(1+\xi)+\frac{1}{2}+\frac{8}{\xi}-\frac{1}{2(1+\xi)^2},
\end{eqnarray}
and
\begin{eqnarray}
\xi=\frac{4E_0\omega_{0}}{m^{2}_{e}}.
\end{eqnarray}
$E_0$ and $\omega_0$ are the incident electron and laser light
energies, and $x=\omega/E_0$. The energy $\omega$ of the scattered
photon depends on its angle $ \theta $ with respect to the
incident electron beam and is given by
\begin{eqnarray}
\omega=\frac{E_{0}(\frac{\xi}{1+\xi})}{1+(\frac{\theta}{\theta_{0}})^{2}}.
\end{eqnarray}
Therefore, at $\theta =0,~\omega=E_{0}\xi/(1+\xi)=\omega_{max}$ is
the maximum energy of the backscattered photon, and
 $x_{max}=\frac{\omega_{max}}{E_{0}}=\frac{\xi}{1+\xi}$.

To avoid unwanted $e^+e^-$ pair production from the collision
between the incident and back-scattered photons, we should not
choose too large $\omega_0$. The threshold for $e^+e^-$ pair
creation is $\omega_{max}\omega_{0} > m^{2}_{e}$, so we require
$\omega_{max}\omega_{0} \leq m^{2}_{e}$. Solving
$\omega_{max}\omega_{0} = m^{2}_{e}$, we find
\begin{eqnarray}
\xi=2(1+\sqrt{2})=4.8.
\end{eqnarray}
For the choice $\xi=4.8,$ we obtain $x_{max}=0.83$ and
$D(\xi_{max})=1.8.$

In the above we have ignored the possible polarization for the photon and
electron beams and we also assume that the number of the
backscattered photons produced per electron is one.

\section{The numerical results and discussions}

In our numerical calculations, the charge conjugate $\bar{t}c$
production channel has also been included. To obtain the numerical
results, we take the SM parameters as $m_{t}=$171.2 GeV,
$m_{c}=$1.25 GeV, $s^{2}_{W}=$0.231, $M_{Z}=$91.2 GeV,
$\alpha_e=1/128$. Moreover, the LHT model has several free
parameters which are related to our study. They are the breaking
scale $f$, 6
parameters($\theta^d_{12},~\theta^d_{13},~\theta^d_{23},~\delta^d_{12},~\delta^d_{13},~\delta^d_{23}$)
in the mixing matrix $V_{H_u}$ and $V_{H_d}$, and the masses of the
mirror quarks. For the mirror quark masses, we get
$m^u_{H_i}=m^d_{H_i}=m_{H_i}(i=1,2,3)$ at $\mathcal {O}(v/f)$ from
Eq.(13). For the matrices $V_{H_u}$ and $V_{H_d}$, considering the
regions of parameter space that only loosely constraint the mass
spectrum of the mirror fermions\cite{FC-LHT3}, we choose two
scenarios as in Ref.\cite{tc-LHT}.

\hspace{1cm} Case I: $V_{H_d}=1,~~~$$V_{H_u}=V^{\dag}_{CKM}$,

\hspace{1cm} Case II:
$s^d_{23}=1/\sqrt{2},~~s^d_{12}=s^d_{13}=0,~~\delta^d_{12}=\delta^d_{23}=\delta^d_{13}=0.$

In both cases, the constraints on the mass spectrum of the mirror
fermions are very relaxed. On the other hand, Ref.\cite{massbound}
has shown that the experimental bounds on four-fermi interactions
involving SM fields provide an upper bound on the mirror fermion
masses and this yields $m_{H_i}\leq4.8f^{2}$. We also consider such
constraint in our calculation. For the breaking scale $f$, we take
two typical values: 500 GeV and 1000 GeV.

For the c.m. energies of the ILC, we choose $\sqrt{s}=500,~1000$ GeV
as examples. Taking account of the detector acceptance, we have
taken the basic cuts on the transverse momentum($p_{T}$) and the
pseudo-rapidity($\eta$) for the final state particles
\begin{eqnarray*}
p_{T}\geq20 GeV, \hspace{1cm}|\eta|\leq 2.5.
\end{eqnarray*}

The numerical results of the cross sections are summarized in
Figs.3-5. Figs.3 and 4 show the cross sections of the processes
$e^+e^-(\gamma\gamma)\rightarrow t\bar{c}$ and $e^-\gamma\rightarrow
e^{-}t\bar{c}$ as a function of $m_{H_3}$ for Case I and Case II,
respectively.
\begin{figure}[h]
\includegraphics {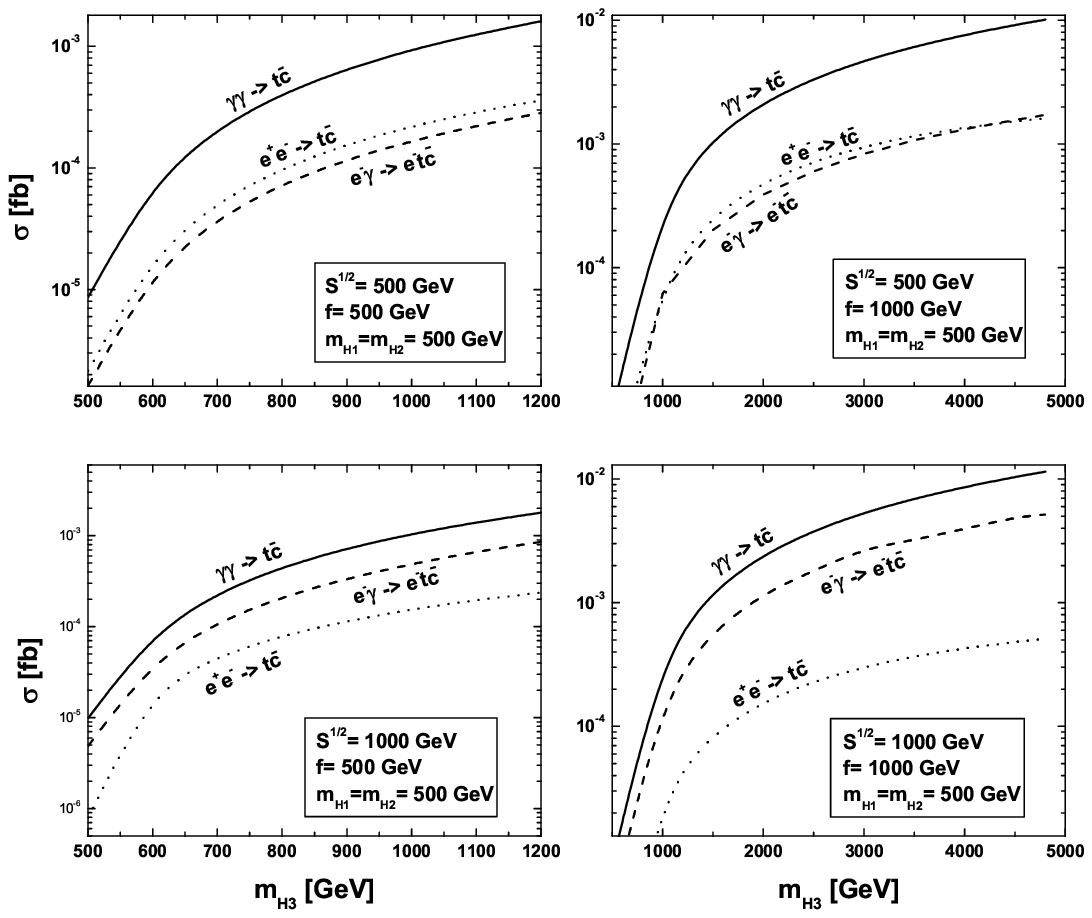}
\caption{\small The cross sections of top-charm associated
production processes versus $m_{H_3}$ in the LHT model for Case I.}
\end{figure}

In Case I, due to the absence of the mixing in the down type gauge
and Goldstone boson interactions, there are no constraints on the
masses of the mirror quarks at one loop-level from the $K$ and $B$
systems and the constraints come only from the $D$ system. The
constraints on the mass of the third generation mirror quark are
very weak. Considering the constraint $m_{H_i}\leq4.8f^{2}$, we take
$m_{H_3}$ to vary in the range of 500-1200 GeV for $f$=500 GeV and
500-4800 GeV for $f$=1000 GeV, and fix $m_{H_1}=m_{H_2}$=500 GeV.

As shown in Fig.3, the cross sections of the three different
production processes rise with the increase of $m_{H_3}$. The reason
is that the couplings between the mirror quarks and the SM quarks
are proportional to the masses of the mirror quarks. The masses of
the heavy gauge bosons and the mirror quarks, $M_{V_{H}}$ and
$m_{H_i}$, are proportional to $f$, but the scale $f$ is insensitive
to the cross sections of these processes because the production
amplitudes are represented in the form of $m_{H_i}/M_{V_{H}}$ which
cancels the effect of $f$. For the case of $\sqrt{s}=1000$ GeV, our
calculations show this case has the slightly larger effects relative
to the case of $\sqrt{s}=500$ GeV.

For Case II, the dependence of the cross sections on $m_{H_3}$ is
presented in Fig.4. In this case, the constraints from the $K$ and B
systems are also very weak. Compared to Case I, the mixing between
the second and third generations is enhanced with the choice of a
bigger mixing angle $s^d_{23}$. Here, we take the same values of
$\sqrt{s}$, $f$ and $m_{H_i}$ as in Case I. Even with stricter
constraints on the masses of the mirror quarks, the large masses of
the mirror quarks can also enhance the cross sections significantly.
The dependence of the cross sections on the c.m. energy is similar
to that in Case I.
\begin{figure}[h]
\includegraphics {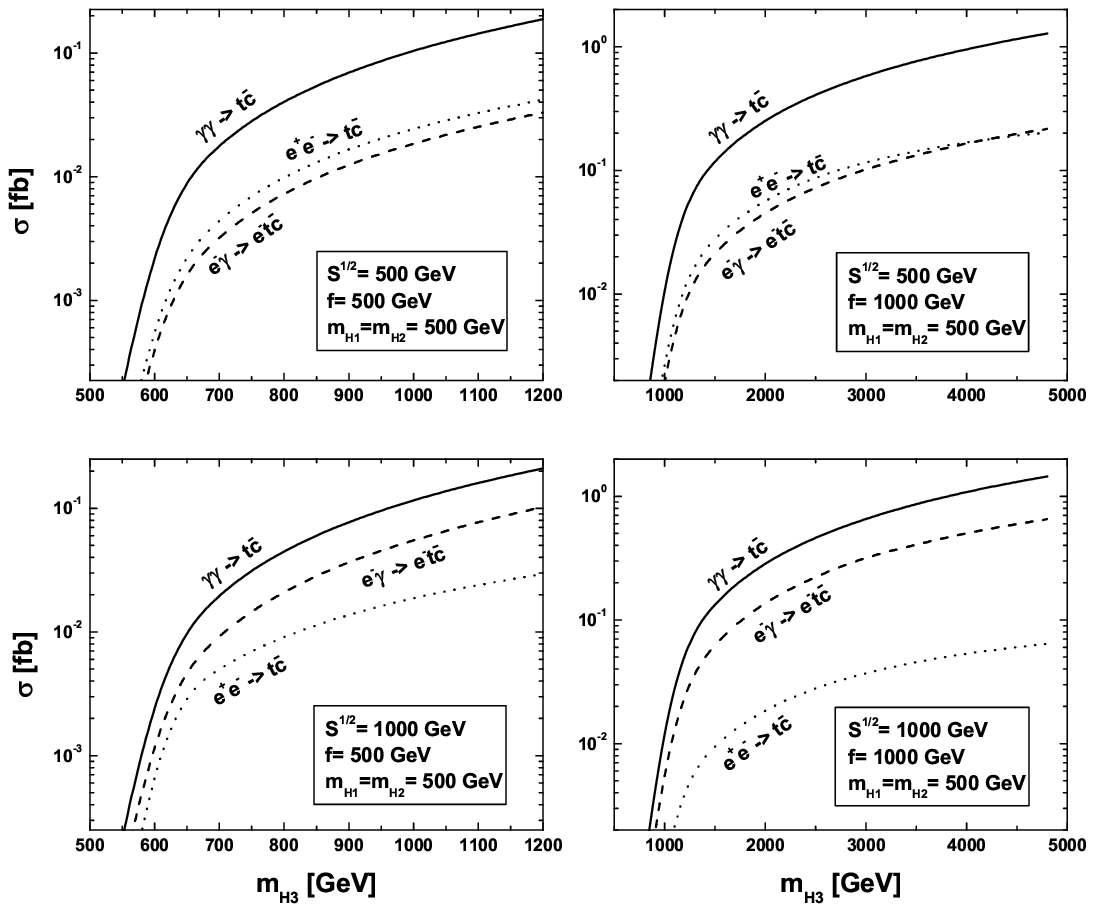}
\caption{\small The cross sections of top-charm associated
production processes versus $m_{H_3}$ in the LHT model for Case II.}
\end{figure}

Among the three processes from Case I and case II, we find that the
cross section of process $\gamma\gamma\rightarrow t\bar{c}$ is the
largest with $\sqrt{s},f=1000$ GeV and heavy mirror quarks. The
optimum value of $\sigma(\gamma\gamma\rightarrow t\bar{c})$ can
reach $\mathcal{O}(10^{0})$ fb. On the other hand, the maximal value
of cross section for process $e^{-}\gamma\rightarrow e^{-}t\bar{c}$
can reach 0.7 fb, which is higher than that in some models such as
MSSM model and type III two Higgs doublet models , but lower than
that in TC2 model\cite{eerrtc-TC2}.

In Fig.5 we show the behavior of the cross sections for
$\gamma\gamma\rightarrow t\bar{c}$, $e^{-}\gamma\rightarrow
e^{-}t\bar{c}$ and $e^+e^-\rightarrow t\bar{c}$ versus the collider
energy for two Cases. We see that the cross section of
$e^+e^-\rightarrow t\bar{c}$ drops quickly with the increase of
collider energy. This is because that the contributions of the LHT
model come from s-channel, so the large c.m. energy $\sqrt{s}$
depresses the cross section. However, for the process
$\gamma\gamma\rightarrow t\bar{c}$, there is only t-channel
contributions, so the large c.m. energy can enhance the cross
section.
\begin{figure}[h]
\includegraphics {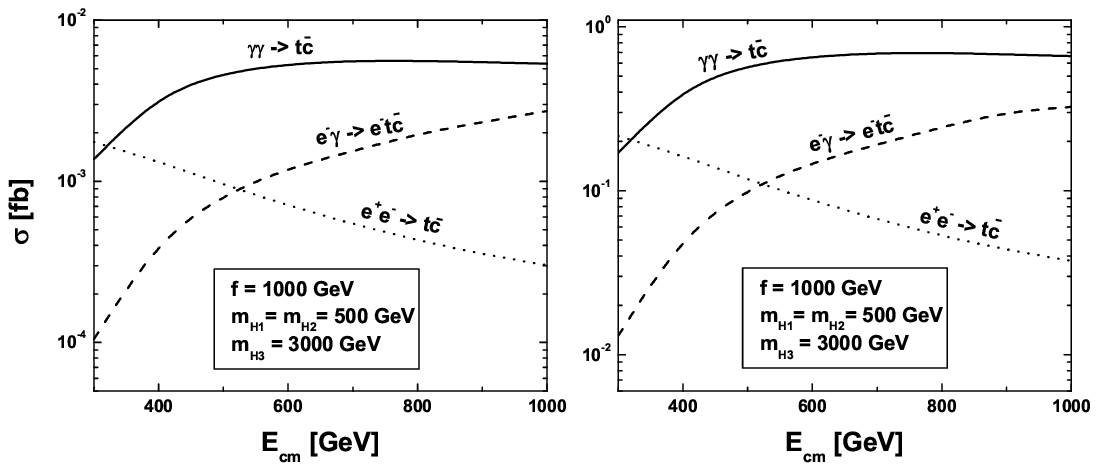}
\caption{\small Cross sections versus collider energy
$E_{cm}=\sqrt{s_{ee}}$ in the LHT model. The left diagram is for
Case I and the right diagram is for Case II.}
\end{figure}

In practice, if we assume conservatively that the signal is reduced
to $10\%$ to eliminate backgrounds, we may expect that the
production $\gamma\gamma\rightarrow t\bar{c}$ as large as $5 fb$ may
be accessible at the ILC at the $3\sigma$ level. According to the
ILC Reference Design Report \cite{ILC}, the total luminosity is
required to be around $500 fb^{-1}$ within the first four years and
about $1000 fb^{-1}$ during the first phase of operation. It means
that the increment of mirror quark masses can sharply enhance the
cross sections of $\gamma\gamma\rightarrow t\bar{c}$ and
$e^{-}\gamma \rightarrow e^{-}t\bar{c}$ to the accessible level at
the ILC. On the other hand, the precise measurement of the cross
sections can certainly provide some information about mirror quark
masses.

Now we discuss the potential to distinguish the different new
physics models via the top-charm production at the ILC. The maximal
values of the cross sections of the processes $\gamma\gamma
\rightarrow t\bar{c}$, $e^{-}\gamma \rightarrow e^{-}t\bar{c}$ and
$e^+e^-\rightarrow t\bar{c}$ in various models are shown in Table 1.

\begin{center}
{\small Table 1: The maximal values of the cross sections of the processes}\\
{\small $\gamma\gamma \rightarrow t\bar{c}$, $e^{-}\gamma \rightarrow e^{-}t\bar{c}$ and $e^+e^-\rightarrow t\bar{c}$ in various models} (\small in fb)\\
 \doublerulesep 0.8pt \tabcolsep 0.05in
\begin{tabular}{|c|c|c|c|c|c|c|}\hline
&SM&2HDM-III &MSSM&TC2&LHT\\
$\gamma\gamma \rightarrow t\bar{c}$ &$\mathcal{O}(10^{-8})$\cite{eerrtc-MSSM} &$\mathcal{O}(10^{-1})$ \cite{rrtc-2HDM}&$\mathcal{O}(10^{-1})$\cite{eerrtc-MSSM}&$\mathcal{O}(10)$\cite{eerrtc-TC2}&$\mathcal{O}(10^{0})$  \\
 \hline
$e^{-}\gamma \rightarrow e^{-}t\bar{c}$ &$\mathcal{O}(10^{-9})$\cite{eerrtc-MSSM} &$\mathcal{O}(10^{-2})$ \cite{eerrtc-TC2}&$\mathcal{O}(10^{-2})$\cite{eerrtc-MSSM}&$\mathcal{O}(1)$\cite{eerrtc-TC2}&$\mathcal{O}(10^{-1})$  \\
 \hline
$e^+e^-\rightarrow t\bar{c}$ &$\mathcal{O}(10^{-10})$\cite{tc-SM}    &$\mathcal{O}(10^{-3})$\cite{eetc-2HDM} &$\mathcal{O}(10^{-2})$\cite{eerrtc-MSSM}&$\mathcal{O}(10^{-1})$\cite{eetc-TC2}&$\mathcal{O}(10^{-2})$  \\
\hline
\end{tabular}
\end{center}
From Table 1 we see that the new physics models can enhance the
SM rates of the FCNC top-charm production processes by several
orders because the tree-level FCNC is absent in the SM. The relation
of the cross section is $\sigma(\gamma\gamma \rightarrow
t\bar{c})>\sigma(e^{-}\gamma \rightarrow
e^{-}t\bar{c})>\sigma(e^{+}e^{-} \rightarrow t\bar{c})$ in every new
physics model. Due to the different values of the cross sections, so
the top-charm production also provides a good way to distinguish the
LHT model from other new physics models.

\section{Conclusion}
We studied the top-charm associated productions via $e^+e^-$,
$e^-\gamma$ and $\gamma\gamma$ collisions in the framework of the
LHT model at the ILC. The numerical results showed that the cross
sections of the processes increase sharply as the mirror quark
masses increase, and in a large part of the allowed parameter space,
the cross sections of $\gamma\gamma\rightarrow t\bar{c}$ and
$e^-\gamma \rightarrow e^{-}t\bar{c}$ may reach the detectable level
at the ILC. If these processes can be observed, some information
about the FC couplings can be obtained in order to distinguish the
LHT model from other new physics. If these FC processes are not to
be observed, the upper limit on mirror quark masses can then be
given.

\section{Acknowledgments}
Y. J. Zhang thanks Junjie Cao and Jinmin Yang for useful
discussions. This work is supported by the National Natural Science
Foundation of China under Grant Nos. 10775039 and 10975047.

\newpage
\begin{center}
\Large{Appendix: The explicit expressions of the effective $t\bar{c}\gamma(Z)$ couplings}
\end{center}

\vspace{1cm}

The effective $t\bar{c}\gamma(Z)$ couplings
$\Gamma^{\mu}_{t\bar{c}\gamma},~\Gamma^{\mu}_{t\bar{c}Z}$ can be
directly calculated based on Fig.1, and they can be represented in
form of 2-point and 3-point standard functions $B_0,B_1,C_{ij}$. In
our calculations, the higher order $v^2/f^2$ terms in the masses of
new gauge bosons and in the Feynman rules are ignored.
$\Gamma^{\mu}_{t\bar{c}\gamma},~\Gamma^{\mu}_{t\bar{c}Z}$ depend on
the momenta of top quark and anti-charm quark($p_t,p_{\bar{c}}$).
Here $p_t$ and $p_{\bar{c}}$ are both outgoing momenta. The explicit
expressions of them are

\begin{eqnarray*}
\Gamma^{\mu}_{t\bar{c}\gamma}(p_t,p_{\bar{c}})&=&\Gamma^{\mu}_{t\bar{c}\gamma}(\eta^{0})+\Gamma^{\mu}_{t\bar{c}\gamma}(\omega^{0})
+\Gamma^{\mu}_{t\bar{c}\gamma}(\omega^{\pm})+\Gamma^{\mu}_{t\bar{c}\gamma}(A_{H})+\Gamma^{\mu}_{t\bar{c}\gamma}(Z_{H})
+\Gamma^{\mu}_{t\bar{c}\gamma}(W_{H}^{\pm})\\&&+\Gamma^{\mu}_{t\bar{c}\gamma}(W_{H}^{\pm}\omega^{\pm}),
\end{eqnarray*}

\begin{eqnarray*}
\Gamma^{\mu}_{t\bar{c}\gamma}(\eta^{0})=\frac{i}{16\pi^{2}}\frac{eg^{\prime2}}{150M_{A_{H}}^{2}}
(V_{Hu})^*_{it}(V_{Hu})_{ic}(A+B+C)~~~~~~~~~~~~~~~~~~~~~~~~~~~~~~~~~\\
A=\frac{1}{p_{t}^{2}-m_{c}^{2}}[m_{Hi}^{2}(m_{c}^{2}B_{0}^{a}+p_{t}^{2}B_{1}^{a})\gamma^{\mu}P_{L}
+m_{t}m_{c}(m_{Hi}^{2}B_{0}^{a}+p_{t}^{2}B_{1}^{a})\gamma^{\mu}P_{R}~~~\\
+m_{t}(m_{Hi}^{2}B_{0}^{a}+m_{c}^{2}B_{1}^{a})\pslash_t\gamma^{\mu}P_{L}
+m_{c}m_{Hi}^{2}(B_{0}^{a}+B_{1}^{a})\pslash_t\gamma^{\mu}P_{R}]~~~\\
B=\frac{1}{p_{\bar{c}}^{2}-m_{t}^{2}}[m_{Hi}^{2}(m_{t}^{2}B_{0}^{b}+p_{\bar{c}}^{2}B_{1}^{b})\gamma^{\mu}P_{L}
+m_{t}m_{c}(m_{Hi}^{2}B_{0}^{b}+p_{\bar{c}}^{2}B_{1}^{b})\gamma^{\mu}P_{R}~~~~\\
-m_{t}m_{Hi}^{2}(B_{0}^{b}+B_{1}^{b})\gamma^{\mu}\pslash_{\bar{c}}P_{L}
-m_{c}(m_{Hi}^{2}B_{0}^{b}+m_{t}^{2}B_{1}^{b})\gamma^{\mu}\pslash_{\bar{c}}P_{R}]~~~\\
C=m_{Hi}^{2}[-\gamma^{\alpha}\gamma^{\mu}\gamma^{\beta}C_{\alpha\beta}^{a}
+\gamma^{\alpha}\gamma^{\mu}(\pslash_{t}+\pslash_{\bar{c}})C_{\alpha}^{a}+2m_{t}C_{\mu}^{a}~~~~~~~~~~~~~~~~~~~~~~~~~~\\
-m_{t}\gamma^{\mu}(\pslash_{t}+\pslash_{\bar{c}})C_{0}^{a}-m_{Hi}^{2}\gamma^{\mu}C_{0}^{a}]
P_{L}~~~~~~~~~~~~~~~~~~~~~~~~~~~~~~~~~~~~\\
+m_{c}[-m_{t}\gamma^{\alpha}\gamma^{\mu}\gamma^{\beta}C_{\alpha\beta}^{a}
+m_{t}\gamma^{\alpha}\gamma^{\mu}(\pslash_{t}+\pslash_{\bar{c}})C_{\alpha}^{a}+2m_{Hi}^{2}C_{\mu}^{a}~~~~~~~~~~~~~~~~~\\
-m_{Hi}^{2}\gamma^{\mu}(\pslash_{t}+\pslash_{\bar{c}})C_{0}^{a}-m_{t}m_{Hi}^{2}\gamma^{\mu}C_{0}^{a}]
P_{R},~~~~~~~~~~~~~~~~~~~~~~~~~~~~~
\end{eqnarray*}

\begin{eqnarray*}
\Gamma^{\mu}_{t\bar{c}\gamma}(\omega^{0})=\frac{i}{16\pi^{2}}\frac{eg^{2}}{6M_{Z_{H}}^{2}}
(V_{Hu})^*_{it}(V_{Hu})_{ic}(D+E+F)~~~~~~~~~~~~~~~~~~~~~~~~~~~~~~~~~\\
D=A(B_{0}^{a}\rightarrow\ B_{0}^{c},B_{1}^{a}\rightarrow\
B_{1}^{c}),~~~~~~~~~~~~~~~~~~~~~~~~~~~~~~~~~~~~~~~~~~~~~~~~~~\\
E=B(B_{0}^{b}\rightarrow\ B_{0}^{d},B_{1}^{b}\rightarrow\
B_{1}^{d}),~~~~~~~~~~~~~~~~~~~~~~~~~~~~~~~~~~~~~~~~~~~~~~~~~~\\
F=C(C_{\alpha\beta}^{a}\rightarrow\
C_{\alpha\beta}^{b},C_{\alpha}^{a}\rightarrow\
C_{\alpha}^{b},C_{0}^{a}\rightarrow\
C_{0}^{b}),~~~~~~~~~~~~~~~~~~~~~~~~~~~~~~~~~
\end{eqnarray*}

\begin{eqnarray*}
\Gamma^{\mu}_{t\bar{c}\gamma}(\omega^{\pm})=\frac{i}{16\pi^{2}}\frac{eg^{2}}{2M_{W_{H}}^{2}}
(V_{Hu})^*_{it}(V_{Hu})_{ic}(\frac{2}{3}G+\frac{2}{3}H-\frac{1}{3}I+J)~~~~~~~~~~~~~~~~~~~~~\\
G=A(B_{0}^{a}\rightarrow\ B_{0}^{e},B_{1}^{a}\rightarrow\
B_{1}^{e}),~~~~~~~~~~~~~~~~~~~~~~~~~~~~~~~~~~~~~~~~~~~~~~~~~~\\
H=B(B_{0}^{b}\rightarrow\ B_{0}^{f},B_{1}^{b}\rightarrow\
B_{1}^{f}),~~~~~~~~~~~~~~~~~~~~~~~~~~~~~~~~~~~~~~~~~~~~~~~~~~\\
I=C(C_{\alpha\beta}^{a}\rightarrow\
C_{\alpha\beta}^{c},C_{\alpha}^{a}\rightarrow\
C_{\alpha}^{c},C_{0}^{a}\rightarrow\
C_{0}^{c}),~~~~~~~~~~~~~~~~~~~~~~~~~~~~~~~~~\\
J=m_{Hi}^{2}[2\gamma^{\alpha}C_{\mu\alpha}^{d}-2(m_{t}-\pslash_{t})C_{\mu}^{d}
+(P_{t}+P_{\bar{c}})^{\mu}\gamma^{\alpha}C_{\alpha}^{d}~~~~~~~~~~~~~~~~~~~~~~~\\
-(P_{t}+P_{\bar{c}})^{\mu}(m_{t}-\pslash_{t})C_{0}^{d}]P_{L}~~~~~~~~~~~~~~~~~~~~~~~~~~~~~~~~~~~~~~~~~~\\
+m_{c}[2m_{t}\gamma^{\alpha}C_{\mu\alpha}^{d}-2(m_{Hi}^{2}-m_{t}\pslash_{t})C_{\mu}^{d}
+m_{t}(P_{t}+P_{\bar{c}})^{\mu}\gamma^{\alpha}C_{\alpha}^{d}~~~~~~~~~~~\\
-(P_{t}+P_{\bar{c}})^{\mu}(m_{Hi}^{2}-m_{t}\pslash_{t})C_{0}^{d}]P_{R},~~~~~~~~~~~~~~~~~~~~~~~~~~~~~~~~~~~~~
\end{eqnarray*}

\begin{eqnarray*}
\Gamma^{\mu}_{t\bar{c}\gamma}(A_{H})=\frac{i}{16\pi^{2}}\frac{eg^{\prime2}}{75}
(V_{Hu})^*_{it}(V_{Hu})_{ic}(K+L+M)~~~~~~~~~~~~~~~~~~~~~~~~~~~~~~~~~~\\
K=\frac{1}{p_{t}^{2}-m_{c}^{2}}[p_{t}^{2}
B_{1}^{a}\gamma^{\mu}P_{L}+m_{c}B_{1}^{a}\pslash_t\gamma^{\mu}P_{R}
]~~~~~~~~~~~~~~~~~~~~~~~~~~~~~~~~~~~~\\
L=\frac{1}{p_{\bar{c}}^{2}-m_{t}^{2}}[p_{\bar{c}}^{2}
B_{1}^{b}\gamma^{\mu}P_{L}-m_{t}B_{1}^{b}\gamma^{\mu}\pslash_{\bar{c}}P_{L}
]~~~~~~~~~~~~~~~~~~~~~~~~~~~~~~~~~~~~\\
M=[-\gamma^{\alpha}\gamma^{\mu}\gamma^{\beta}C_{\alpha\beta}^{a}
+(\pslash_{t}+\pslash_{\bar{c}})\gamma^{\mu}\gamma^{\alpha}C_{\alpha}^{a}
-m_{Hi}^{2}\gamma^{\mu}C_{0}^{a}] P_{L},~~~~~~~~~~~~~~~~~~~
\end{eqnarray*}

\begin{eqnarray*}
\Gamma^{\mu}_{t\bar{c}\gamma}(Z_{H})=\frac{i}{16\pi^{2}}\frac{eg^{2}}{3}
(V_{Hu})^*_{it}(V_{Hu})_{ic}(N+O+P)~~~~~~~~~~~~~~~~~~~~~~~~~~~~~~~~~~~\\
N=K(B_{1}^{a}\rightarrow\ B_{1}^{c}),~~~~~~~~~~~~~~~~~~~~~~~~~~
~~~~~~~~~~~~~~~~~~~~~~~~~~~~~~~~~~~~\\
O=L(B_{1}^{b}\rightarrow\ B_{1}^{d}),~~~~~~~~~~~~~~~~~~~~~~~~~~~~
~~~~~~~~~~~~~~~~~~~~~~~~~~~~~~~~~~~\\
P=M(C_{\alpha\beta}^{a}\rightarrow\
C_{\alpha\beta}^{b},C_{\alpha}^{a}\rightarrow\
C_{\alpha}^{b},C_{0}^{a}\rightarrow\
C_{0}^{b}),~~~~~~~~~~~~~~~~~~~~~~~~~~~~~~~
\end{eqnarray*}
\begin{eqnarray*}
\Gamma^{\mu}_{t\bar{c}\gamma}(W_{H}^{\pm})=\frac{i}{16\pi^{2}}\frac{eg^{2}}{2}
(V_{Hu})^*_{it}(V_{Hu})_{ic}(\frac{4}{3}Q+\frac{4}{3}R-\frac{2}{3}S-T)~~~~~~~~~~~~~~~~~~~~~~~~~\\
Q=K(B_{1}^{a}\rightarrow\ B_{1}^{e}),~~~~~~~~~~~~~~~~~~~~~~~~
~~~~~~~~~~~~~~~~~~~~~~~~~~~~~~~~~~~~~~~~\\
R=L(B_{1}^{b}\rightarrow\
B_{1}^{f}),~~~~~~~~~~~~~~~~~~~~~~~~~~~~~~~~
~~~~~~~~~~~~~~~~~~~~~~~~~~~~~~~~~\\
S=M(C_{\alpha\beta}^{a}\rightarrow\
C_{\alpha\beta}^{c},C_{\alpha}^{a}\rightarrow\
C_{\alpha}^{c},C_{0}^{a}\rightarrow\ C_{0}^{c})~~~~~~~~~~~~~~~~~~~~~~~~~~~~~~~~~~\\
T=\{4\gamma^{\alpha}C_{\mu\alpha}^{d}+[\gamma^{\mu}\pslash_{t}\gamma^{\alpha}
-\gamma^{\mu}\gamma^{\alpha}\pslash_{t}+2\pslash_{t}\gamma^{\alpha}\gamma^{\mu}
+\gamma^{\alpha}\pslash_{t}\gamma^{\mu}-\gamma^{\mu}\gamma^{\alpha}\pslash_{\bar{c}}~~~~~~~~\\
+2\pslash_{\bar{c}}\gamma^{\alpha}\gamma^{\mu}
+2(p_{t}+p_{\bar{c}})^{\mu}\gamma^{\alpha}]C_{\alpha}^{d}+4\pslash_{t}C_{\mu}^{d}
+2(B_{0}^{g}+m_{W_{H}}^{2}C_{0}^{d})\gamma^{\mu}~~~~~~\\+[2\pslash_{\bar{c}}\pslash_{t}\gamma^{\mu}
-\gamma^{\mu}\pslash_{t}\pslash_{\bar{c}}
+2\pslash_{t}(p_{t}+p_{\bar{c}})^{\mu}
+p_{t}^{2}\gamma^{\mu}]C_{0}^{d}\}P_{L},~~~~~~~~~~~~~~~~~~
\end{eqnarray*}

\begin{eqnarray*}
\Gamma^{\mu}_{t\bar{c}\gamma}(W_{H}^{\pm}\omega^{\pm})=\frac{i}{16\pi^{2}}\frac{eg^{2}}{2}
(V_{Hu})^*_{it}(V_{Hu})_{ic}~~~~~~~~~~~~~~~~~~~~~~~~
~~~~~~~~~~~~~~~~~~~~~~~~~~\\
\times\{[m_{t}\pslash_{t}C_{0}^{d}+m_{t}\gamma^{\alpha}C_{\alpha}^{d}
-2m_{Hi}^{2}C_{0}^{d}]\gamma^{\mu}P_{L}
+m_{c}[\gamma^{\mu}\pslash_{t}C_{0}^{d}+\gamma^{\mu}\gamma^{\alpha}C_{\alpha}^{d}]P_{R}\},
\end{eqnarray*}

\begin{eqnarray*}
\Gamma^{\mu}_{t\bar{c}Z}(p_t,p_{\bar{c}})=\Gamma^{\mu}_{t\bar{c}Z}(\eta^{0})
+\Gamma^{\mu}_{t\bar{c}Z}(\omega^{0})
+\Gamma^{\mu}_{t\bar{c}Z}(\omega^{\pm})+\Gamma^{\mu}_{t\bar{c}Z}(A_{H})+\Gamma^{\mu}_{t\bar{c}Z}(Z_{H})
+\Gamma^{\mu}_{t\bar{c}Z}(W_{H}^{\pm})\\
+\Gamma^{\mu}_{t\bar{c}Z}(W_{H}^{\pm}\omega^{\pm}),~~~~~~~~~~~~~~~~~~~~~~~~~~~~~~~~~~~~~~~~~~~~~~~~~~~~~~
~~~~~~~~~~~~~
\end{eqnarray*}
\begin{eqnarray*}
\Gamma^{\mu}_{t\bar{c}Z}(\eta^{0})=\frac{i}{16\pi^{2}}\frac{g}{\cos\theta_{W}}
\frac{g^{\prime2}}{100M_{A_{H}}^{2}}(V_{Hu})^*_{it}(V_{Hu})_{ic}(A'+B'+C')~~~~~~~~~~~~~~~~~~~~\\
A'=\frac{1}{p_{t}^{2}-m_{c}^{2}}[
(\frac{1}{2}-\frac{2}{3}\sin^{2}\theta_{W})m_{Hi}^{2}(m_{c}^{2}B_{0}^{a}+p_{t}^{2}B_{1}^{a})\gamma^{\mu}P_{L}~~~~~~~~~~~~~~~~~~~~\\
-\frac{2}{3}\sin^{2}\theta_{W}m_{t}m_{c}(m_{Hi}^{2}B_{0}^{a}+p_{t}^{2}B_{1}^{a})\gamma^{\mu}P_{R}~~~~~~~~~~~~~~~~~~~~~~\\
+(\frac{1}{2}-\frac{2}{3}\sin^{2}\theta_{W})m_{t}(m_{Hi}^{2}B_{0}^{a}+m_{c}^{2}B_{1}^{a})\pslash_t\gamma^{\mu}P_{L}~~~~~~~~~~~~~~\\
-\frac{2}{3}\sin^{2}\theta_{W}m_{c}m_{Hi}^{2}(B_{0}^{a}+B_{1}^{a})\pslash_t\gamma^{\mu}P_{R}]~~~~~~~~~~~~~~~~~~~~~~~~~\\
B'=\frac{1}{p_{\bar{c}}^{2}-m_{t}^{2}}[(\frac{1}{2}-\frac{2}{3}\sin^{2}\theta_{W})m_{Hi}^{2}(m_{t}^{2}B_{0}^{b}
+p_{\bar{c}}^{2}B_{1}^{b})\gamma^{\mu}P_{L}~~~~~~~~~~~~~~~~~~~~\\
-\frac{2}{3}\sin^{2}\theta_{W}m_{t}m_{c}(m_{Hi}^{2}B_{0}^{b}+p_{\bar{c}}^{2}B_{1}^{b})\gamma^{\mu}P_{R}~~~~~~~~~~~~~~~~~~~~~~\\
+\frac{2}{3}\sin^{2}\theta_{W}m_{t}m_{Hi}^{2}(B_{0}^{b}+B_{1}^{b})\gamma^{\mu}\pslash_{\bar{c}}P_{L}~~~~~~~~~~~~~~~~~~~~~~~~~~\\
-(\frac{1}{2}-\frac{2}{3}\sin^{2}\theta_{W})m_{c}(m_{Hi}^{2}B_{0}^{b}+m_{t}^{2}B_{1}^{b})\gamma^{\mu}\pslash_{\bar{c}}P_{R}]~~~~~~~~~~~~~\\
C'=(\frac{1}{2}-\frac{2}{3}\sin^{2}\theta_{W})C,~~~~~~~~~~~~~~~~~~~~~~~~~~~~~~~~~~~~~~~~~~~~~~~~~~~~~~~~~~~~
\end{eqnarray*}

\begin{eqnarray*}
\Gamma^{\mu}_{t\bar{c}Z}(\omega^{0})=\frac{i}{16\pi^{2}}\frac{g}{\cos\theta_{W}}
\frac{g^{2}}{4M_{Z_{H}}^{2}}
(V_{Hu})^*_{it}(V_{Hu})_{ic}(D'+E'+F')~~~~~~~~~~~~~~~~~~~~~~~~~\\
D'=A'(B_{0}^{a}\rightarrow\ B_{0}^{c},B_{1}^{a}\rightarrow\
B_{1}^{c}),~~~~~~~~~~~~~~~~~~~~~~~~~~~~~~~~~~~~~~~~~~~~~~~~~~~~\\
E'=B'(B_{0}^{b}\rightarrow\
B_{0}^{d},B_{1}^{b}\rightarrow\ B_{1}^{d}),~~~~~~~~~~~~~~~~~~~~~~~~~~~~~~~~~~~~~~~~~~~~~~~~~~~~\\
F'=C'(C_{\alpha\beta}^{a}\rightarrow\
C_{\alpha\beta}^{b},C_{\alpha}^{a}\rightarrow\
C_{\alpha}^{b},C_{0}^{a}\rightarrow\
C_{0}^{b}),~~~~~~~~~~~~~~~~~~~~~~~~~~~~~~~~~~~
\end{eqnarray*}

\begin{eqnarray*}
\Gamma^{\mu}_{t\bar{c}Z}(\omega^{\pm})=\frac{i}{16\pi^{2}}\frac{g}{\cos\theta_{W}}
\frac{g^{2}}{2M_{W_{H}}^{2}}(V_{Hu})^*_{it}(V_{Hu})_{ic}(G'+H'+I'+J')~~~~~~~~~~~~~~~~~~~\\
G'=A'(B_{0}^{a}\rightarrow\ B_{0}^{e},B_{1}^{a}\rightarrow\
B_{1}^{e}),~~~~~~~~~~~~~~~~~~~~~~~~~~~~~~~~~~~~~~~~~~~~~~~~~~~~~\\
H'=B'(B_{0}^{b}\rightarrow\
B_{0}^{f},B_{1}^{b}\rightarrow\ B_{1}^{f}),~~~~~~~~~~~~~~~~~~~~~~~~~~~~~~~~~~~~~~~~~~~~~~~~~~~~~\\
I'=(-\frac{1}{2}+\frac{1}{3}\sin^{2}\theta_{W})I,~~~~~~~~~~~~~~~~~~~~~~~~~~~~~~~~~~~~~~~~~~~~~~~~~~~~~~~~~~~~~\\
J'=cos^{2}\theta_{W}J,~~~~~~~~~~~~~~~~~~~~~~~~~~~~~~~~~~~~~~~~~~~~~~~~~~~~~~~~~~~~~~~~~~~~~~~~~~~\\
\end{eqnarray*}

\begin{eqnarray*}
\Gamma^{\mu}_{t\bar{c}Z}(A_{H})=\frac{i}{16\pi^{2}}\frac{g}{\cos\theta_{W}}
\frac{g^{\prime2}}{50}
(V_{Hu})^*_{it}(V_{Hu})_{ic}(K'+L'+M')~~~~~~~~~~~~~~~~~~~~~~~~~~~~~~~~~\\
K'=\frac{1}{p_{t}^{2}-m_{c}^{2}}[(\frac{1}{2}-\frac{2}{3}\sin^{2}\theta_{W})p_{t}^{2}
B_{1}^{a}\gamma^{\mu}P_{L}-\frac{2}{3}\sin^{2}\theta_{W}m_{c}B_{1}^{a}\pslash_t\gamma^{\mu}P_{R}
]~~~~~~~~~~~~\\
L'=\frac{1}{p_{\bar{c}}^{2}-m_{t}^{2}}[(\frac{1}{2}-\frac{2}{3}\sin^{2}\theta_{W})p_{\bar{c}}^{2}
B_{1}^{b}\gamma^{\mu}P_{L}+\frac{2}{3}\sin^{2}\theta_{W}m_{t}B_{1}^{b}\gamma^{\mu}\pslash_{\bar{c}}P_{L}
]~~~~~~~~~~~~~\\
M'=(\frac{1}{2}-\frac{2}{3}\sin^{2}\theta_{W})[-\gamma^{\alpha}\gamma^{\mu}\gamma^{\beta}C_{\alpha\beta}^{a}+
(\pslash_t+\pslash_{\bar{c}})\gamma^{\mu}\gamma^{\alpha}C_{\alpha}^{a}
-m_{Hi}^{2}C_{0}^{a}\gamma^{\mu} ]P_{L},~~~~~~
\end{eqnarray*}

\begin{eqnarray*}
\Gamma^{\mu}_{t\bar{c}Z}(Z_{H})=\frac{i}{16\pi^{2}}\frac{g}{\cos\theta_{W}}
\frac{g^{2}}{2}(V_{Hu})^*_{it}(V_{Hu})_{ic}(N'+O'+P')~~~~~~~~~~~~~~~~~~~~~~~~~~~~~~~~~\\
N'=K'(B_{1}^{a}\rightarrow\ B_{1}^{c}),~~~~~~~~~~~~~~~~~~~~~~~~~~~~~~~~~~~~~~~~~~~~~~~~~~~~~~~~~~~~~~~~~~~~~~\\
O'=L'(B_{1}^{b}\rightarrow\ B_{1}^{d}),~~~~~~~~~~~~~~~~~~~~~~~~~~~~~~~~~~~~~~~~~~~~~~~~~~~~~~~~~~~~~~~~~~~~~~~\\
P'=M'(C_{\alpha\beta}^{a}\rightarrow\
C_{\alpha\beta}^{b},C_{\alpha}^{a}\rightarrow\
C_{\alpha}^{b},C_{0}^{a}\rightarrow\
C_{0}^{b}),~~~~~~~~~~~~~~~~~~~~~~~~~~~~~~~~~~~~~~
\end{eqnarray*}
\begin{eqnarray*}
\Gamma^{\mu}_{t\bar{c}Z}(W_{H}^{\pm})=\frac{i}{16\pi^{2}}\frac{g}{\cos\theta_{W}}
{g^{2}}(V_{Hu})^*_{it}(V_{Hu})_{ic}(Q'+R'+S'+T')~~~~~~~~~~~~~~~~~~~~~~~~~~~\\
Q'=K'(B_{1}^{a}\rightarrow\ B_{1}^{e}),~~~~~~~~~~~~~~~~~~~~~~~~~~~~~~~~~~~~~~~~~~~~~~~~~~~~~~~~~~~~~~~~~~~~~\\
R'=L'(B_{1}^{b}\rightarrow\ B_{1}^{f}),~~~~~~~~~~~~~~~~~~~~~~~~~~~~~~~~~~~~~~~~~~~~~~~~~~~~~~~~~~~~~~~~~~~~~~\\
S'=(-\frac{1}{2}+\frac{1}{3}\sin^{2}\theta_{W})S,~~~~~~~~~~~~~~~~~~~~~~~~~~~~~~~~~~~~~~~~~~~~~~~~~~~~~~~~~~~~~~~~\\
T'=-\frac{1}{2}\cos^{2}\theta_{W}T,~~~~~~~~~~~~~~~~~~~~~~~~~~~~~~~~~~~~~~~~~~~~~~~~~~~~~~~~~~~~~~~~~~~~~~~
\end{eqnarray*}
\begin{eqnarray*}
\Gamma^{\mu}_{t\bar{c}Z}(W_{H}^{\pm}\omega^{\pm})=\frac{i}{16\pi^{2}}g\cos\theta_{W}\frac{g^{2}}{2}
(V_{Hu})^*_{it}(V_{Hu})_{ic}~~~~~~~~~~~~~~~~~~~~~~~~~~~~~~~~~~~~~~~~~~~~~~~~\\
\times\{[m_{t}\pslash_{t}C_{0}^{d}+m_{t}\gamma^{\alpha}C_{\alpha}^{d}-2m_{Hi}^{2}C_{0}^{d}]\gamma^{\mu}P_{L}
+m_{c}[\gamma^{\mu}\pslash_{t}C_{0}^{d}+\gamma^{\mu}\gamma^{\alpha}C_{\alpha}^{d}]P_{R}\}.
\end{eqnarray*}
For the two-point and three-point standard loop functions
$B_0,B_1,C_0,~C_{ij}$ in the above expressions are defined as
\begin{eqnarray*}
B^{a}=B^{a}(-p_{t},m_{Hi},M_{A_{H}}),
B^{b}=B^{b}(-p_{\bar{c}},M_{Hi},M_{A_{H}}),\\
B^{c}=B^{c}(-p_{t},m_{Hi},M_{Z_{H}}),
B^{d}=B^{d}(-p_{\bar{c}},M_{Hi},M_{Z_{H}}),\\
B^{e}=B^{e}(-p_{t},m_{Hi},M_{W_{H}}),
B^{f}=B^{f}(-p_{\bar{c}},M_{Hi},M_{W_{H}}),\\
B^{g}=B^{g}(p_{\bar{c}},M_{Hi},M_{W_{H}}),~~~~~~~~~~~~~~~~~~~~~~~~~~~~~~~~~~~~~\\
C_{ij}^{a}=C_{ij}^{a}(-p_{t},-p_{\bar{c}},m_{Hi},M_{A_{H}},m_{Hi}),~~~~~~~~~~~~~~~~~~~~\\
C_{ij}^{b}=C_{ij}^{b}(-p_{t},-p_{\bar{c}},m_{Hi},M_{Z_{H}},m_{Hi}),~~~~~~~~~~~~~~~~~~~~\\
C_{ij}^{c}=C_{ij}^{c}(-p_{t},-p_{\bar{c}},m_{Hi},M_{W_{H}},m_{Hi}),~~~~~~~~~~~~~~~~~~~~\\
C_{ij}^{d}=C_{ij}^{d}(p_{t},p_{\bar{c}},M_{W_{H}},m_{Hi},M_{W_{H}}).~~~~~~~~~~~~~~~~~~~~~~~\\
\end{eqnarray*}

\begin{thebibliography}{99}

\bibitem{little Higgs}
N. Arkani-Hamed, A. G. Cohen and H. Georgi, Phys. Lett. B{\bf 513}, 232 (2001);
N. Arkani-Hamed, et al., JHEP {\bf 0208}, 020 (2002).

\bibitem{LH}
N. Arkani-Hamed, A. G. Cohen, E. Katz, A. E. Nelson, JHEP {\bf 0207}, 034 (2002).

\bibitem{constraints}
J. L. Hewett, F. J. Petriello, and T. G. Rizzo, JHEP {\bf 0310}, 062(2003);
C. Csaki, et al., Phys. Rev. D{\bf 67}, 115002 (2003); Phys. Rev. D{\bf 68}, 035009 (2003);
M. C. Chen, S. Dawson, Phys. Rev. D{\bf 70}, 015003 (2004);
W. Kilian and J. Reuter, Phys. Rev. D{\bf 70}, 015004 (2004).

\bibitem{fine-tuning}
 G. Marandella, C. Schappacher, and A. Strumia,  Phys. Rev. D{\bf 72}, 035014 (2005).

\bibitem{LHT}I. Low, JHEP {\bf 0410}, 067 (2004);
H. C. Cheng and I. Low, JHEP, {\bf 0408}, 061 (2004);
J. Hubisz and P.Meade, Phys. Rev. D{\bf 71}, 035016 (2005);
J. Hubisz, S. J. Lee and G. Paz, JHEP {\bf 0606}, 041 (2006).

\bibitem{scale} J. Hubisz, P. Meade, A. Noble, and M. Perelstein, JHEP {\bf 0601}, 136 (2006);

\bibitem{FC-LHT3} J. Hubisz, S. J. Lee and G. Paz, JHEP {\bf 0606}, 041 (2006).

\bibitem{FC-LHT2} M. Blanke, {\it et al.},  JHEP {\bf 0612}, 003 (2006).

\bibitem{FC-LHT1} M. Blanke, {\it et al.}, JHEP {\bf 0701}, 066 (2007).

\bibitem{FC-LHT6} M. Blanke, {\it et al.}, JHEP {\bf 0706}, 082 (2007).
\bibitem{FC-LHT5} M. Blanke, {\it et al.}, Phys. Lett. B {\bf 657}, 81 (2007).
\bibitem{FC-LHT4} M. Blanke, {\it et al.},  JHEP {\bf 0705}, 013 (2007);
S. R. Choudhury,  {\it et al.}, hep-ph/0612327.

\bibitem{ILC} J. Brau, Y. Okada, and N. Walker, arXiv: 0712.1950;
              A. Djouadi et al., arXiv:0709.1893;
              N. Phinney, N. Toge, and N. Walker, arXiv: 0712.2361;
              T. Behnke, C. damerell, J. Jaros, and A. Myamoto, arXiv:0712.2356.


\bibitem{er} K. Abe et al., ACFA Linear Collider Working Group, hep-ph/0109166.

\bibitem{Rare top decay} For examples, see,
X. L. Wang, {\it et.al.}, Phys. Rev. D{\bf 50}, 5781 (1994);
G. R. Lu, F. R. Yin, X. L. Wang, L. D. Wan, Phys. Rev. D{\bf 68}, 015002(2003);
C. S. Li, R. J. Oakes, J. M. Yang, Phys. Rev. D{\bf 49}, 293 (1994);
G. Couture, C. Hamzaoui, H. Konig, Phys. Rev. D{\bf 52}, 1713 (1995);
J. L. Lopez, D. V. Nanopoulos, R. Rangarajan, Phys. Rev. D{\bf 56}, 3100 (1997);
G. M. de Divitiis, R. Petronzio, L. Silvestrini, Nucl. Phys. B{\bf 504}, 45 (1997);
J. M. Yang, B.-L. Young, X. Zhang, Phys. Rev. D{\bf 58}, 055001 (1998);
J. M. Yang, C. S. Li, Phys. Rev. D{\bf 49}, 3412 (1994);
J. Guasch, J. Sola, Nucl. Phys. B{\bf 562}, 3 (1999);
G. Eilam, J. L. Hewett and A. Soni, Phys. Rev. D{\bf 44}, 1473 (1991);
G. Eilam, {\it et al.}, Phys. Lett. B{\bf 510}, 227 (2001);
 J.~Cao, {\it et al.}, Phys. Rev. D74, 031701 (2006).

\bibitem{eerrtc-MSSM}
J. Cao, Z. Xiong, J. M. Yang, Nucl. Phys. B{\bf 651}, 87 (2003);
C. S. Li, X. Zhang, S. H. Zhu, Phys. Rev. D{\bf 60}, 077702 (1999).

\bibitem{rrtc-MSSM} Z. H. Yu, {\it et.al.},  Eur. Phys. J. C{\bf 16}, 541 (2000).

\bibitem{eetc-2HDM}
D. Atwood, L. Reina and A. Soni, Phys. Rev. D{\bf 53}, 1199 (1996);
S. Bar-Shalom, G. Eilam, A. Soni and J. Wudka, Phys. Rev. Lett. {\bf 79}, 1217 (1997);
                                               Phys. Rev. D{\bf 57}, 2957 (1998);
D. Atwood, L. Reina and A. Soni, Phys. Rev. D{\bf 55}, 3156 (1997);
W.-S. Hou, G.-L. Lin and C.-Y. Ma, Phys. Rev. D{\bf 56}, 7434 (1997).

\bibitem{rrtc-2HDM}
Y. Jiang, {\it et.al.}, Phys. Rev. D{\bf 57}, 4343 (1998);
W. S. Hou and G. L. Lin, Phys. Lett. B{\bf 379}, 261 (1996).

\bibitem{eerrtc-TC2}
J. Cao, G. Liu, J. M. Yang,  Eur. Phys. J. C{\bf 41}, 381 (2005).

\bibitem{eetc-TC2} C. Yue, Y. Dai, Q. Xu, G. Liu, Phys. Lett. B{\bf 525}, 301 (2002).

\bibitem{rrtc-TC2} C. Yue, G. R. Lu, J. Cao, J. Li, G. Liu, Phys. Lett. B{\bf 496}, 93 (2000).

\bibitem{tc-effective-LC1}
T. Han and J. L. Hewett, Phys. Rev. D{\bf 60}, 074015 (1999);
J. A. Aguilar-Saavedra, Phys. Lett. B{\bf 502}, 115 (2001);
J. A. Aguilar-Saavedra, T. Riemann, hep-ph/0102197.

\bibitem{tc-effective-LC2}S. Bar-Shalom and J. Wudka, Phys. Rev. D{\bf 60}, 094016 (1999).

\bibitem{tc-effective-LC3}
V. F. Obraztsov, S. R. Slabospitsky and O. P. Yushchenko, Phys. Lett. B{\bf 426}, 393 (1998);
U. Mahanta and A. Ghosal, Phys. Rev. D{\bf 57}, 1735 (1998).

\bibitem{wang}
X. L. Wang, {\it et.al.}, Phys. Rev. D{\bf 66}, 075009 (2002);
X. L. Wang, B. Z. Li, Y. L. Yang, Phys. Rev. D{\bf 68}, 115003 (2003);
W. N. Xu, X. L. Wang, Z. J. Xiao,  Eur. Phys. J. C{\bf 51} 891 (2007).

\bibitem{tcv-LHT}
H. S. Hou, Phys. Rev. D{\bf 75}, 094010 (2007);
X. L. Wang, Y. J. Zhang, H. L. Jin, Y. H. Xi, Nucl. Phys. B{\bf 810}, 226 (2009);
X. F. Han, L. Wang, and J. M. Yang, arXiv: 0903.5491 [hep-ph].

\bibitem{tc-LHT}
X. L. Wang, H. L. Jin, Y. J. Zhang, Y. H. Xi, Nucl. Phys. B{\bf 807}, 210 (2009).

\bibitem{et-LHT} C. X. Yue, J. Wen, J. Y. Liu, W. Liu, hep-ph/08031335.

\bibitem{tc-SM}
C. S. Huang, X. H. Wu and S. H. Zhu, Phys. Lett. B{\bf 452}, 143 (1999);
C.-H Chang et al., Phys. Lett. B{\bf 313}, 389 (1993);
A. Axelrod, Nucl. Phys. B{\bf 209}, 349 (1982);
M. Clements et al., Phys. Rev. D{\bf 27}, 570 (1983);
V. Ganapathi et al., Phys. Rev. D{\bf 27}, 579 (1983);
G. Eilam, Phys. Rev. D{\bf 28}, 1202 (1983).

\bibitem{method}
 J. J. Cao, G. Eilam, M. Frank, K. Hikasa, G. L. Liu, I. Turan, J. M. Yang, Phys. Rev. D75, 075021 (2007).
\bibitem{distribution}
G. Jikia, Nucl. Phys. B{\bf 374}, 83 (1992);
O. J. P. Eboli, et~al., Phys. Rev. D{\bf 47}, 1889 (1993);
K. M. Cheung, {\em ibid.} {\bf 47}, 3750 (1993).

\bibitem{massbound}
J. Hubisz, P. Meade, A. Noble, M. Perelstein, JHEP {\bf 0601}, 135 (2006).
\end{thebibliography}
\end{document}